\documentclass[a4paper,10pt]{article}
\usepackage{jheppub} 

\usepackage{xcolor}

\usepackage{array}
\usepackage{multirow}
\usepackage{amsmath}
\usepackage{slashed}
\usepackage{IEEEtrantools}
\usepackage{amstext,amssymb}
\usepackage{graphicx}
\usepackage{graphicx,wrapfig}
\usepackage{orcidlink}
\usepackage{hyperref}
\usepackage{comment}

\title{Predictions of Modular Symmetry Fixed Points on Neutrino Masses, Mixing, and Leptogenesis}

\author{Priya$^1$,}
\emailAdd{kashyappriya963@gmail.com}

\author{B. C. Chauhan$^1$,}
\emailAdd{bcawake@hpcu.ac.in}

\author{Deepak Kumar$^{2}$}
\emailAdd{deepakk@iiserbpr.ac.in}

\author{and Takaaki Nomura$^3$}
\emailAdd{nomura@scu.edu.cn}

\affiliation{$^1$Department of Physics and Astronomical Science, Central University of Himachal Pradesh, Dharamshala (HP) 176215, India}
\affiliation{$^2$Department of Physics, Indian Institute of Science Education and Research Berhampur, Odisha, 760003, India}
\affiliation{$^3$College of Physics, Sichuan University, Chengdu 610065, China}

\abstract{
In recently proposed framework of non-holomorphic modular symmetry introduces the concept of negative and zero modular weight of Yukawa couplings. These Yukawa couplings are function of complex modulus $\tau$, which is responsible for the CP asymmetry produced during leptogenesis. In this work, we restrict the $\tau$ on the fixed points of modular symmetry rather than its fundamental domain in such manner Yukawa couplings are also get fixed. We have adopt this framework and propose a type III seesaw mechanism. The model is tested against neutrino oscillation data through a $\chi^2$ analysis using NuFIT~6.1. To test the stability of these predictions, we also analyze regions near each fixed point by introducing a deviation $\tau \rightarrow \tau_{\rm fixed}(1 + \epsilon e^{i\phi})$ with $\epsilon \in (0,0.1)$ and $\phi \in (-\pi,\pi)$. Our results show that certain fixed points, along with their nearby regions, are capable of producing viable neutrino phenomenology while also generating the observed baryon asymmetry of the Universe.}


\makeatletter \gdef\@fpheader{} \makeatother
\begin{document}
\maketitle

\section{Introduction} \label{introduction}
The Standard Model(SM) of particle physics explains all the elementry particles with the neutrinos as massless left-handed Weyl fermions. However, the neutrino oscillation experiments, such as Super kamiokande~\cite{Super-Kamiokande:1998kpq}, Sudbary Neutrino observatory (SNO)~\cite{SNO:2002tuh} and KAmland \cite{KamLAND:2002uet} experiments confirmed that neutrinos are massive. This results provides direct experimental motivation for new physics beyond the Standard Model (BSM). Precision measurements from reactor and accelerator-based neutrino oscillation experiments have established all three mixing angles($\sin^2\theta_{13}$, $\sin^2\theta_{12}$, $\sin^2\theta_{23}$) and mass-squared differences($\Delta m^2_{\rm solar}$ and $\Delta m^2_{\rm atm}$) with high accuracy. Despite this progress, the mechanism responsible for neutrino mass generation and the fundamental nature of neutrinos remain open questions, motivating extensions of the SM. The smallness of neutrino masses can be naturally explained by extending the SM with right-handed neutrinos through the seesaw mechanisms. In particular, the type-I seesaw mechanism arises from the addition of heavy fermionic fields that are singlets under the SM gauge group, i.e. $SU(2)_L$ singlet fermions, which generate light Majorana neutrino masses after electroweak symmetry breaking~\cite{Minkowski:1977sc}. In contrast, the type-II seesaw mechanism is realized by introducing an $SU(2)_L$ scalar triplet with hypercharge $Y=1$, whose vacuum expectation value directly contributes to the neutrino mass matrix~\cite{Mohapatra:1979ia}. The type-III seesaw mechanism involves the addition of fermionic fields transforming as triplets under $SU(2)_L$ with zero hypercharge, which similarly induce small neutrino masses through their heavy Majorana mass terms~\cite{Foot:1988aq}. Neutrino oscillation experiments have firmly established that neutrinos are massive and mix among flavors. After the successful measurement of the reactor mixing angle, all neutrino oscillation parameters have been determined with high precision, and the corresponding global-fit values, as provided by the NuFIT 6.1 analysis~\cite{NuFit61,Esteban:2024eli}, are presented in a Table~\ref{tab:neutrino_data}. Despite significant experimental progress, the fundamental nature of neutrinos remains unresolved, as they may be either Dirac or Majorana fermions. In addition, the observed matter-antimatter asymmetry of the Universe is another open problem that can be addressed within the seesaw framework via leptogenesis, where the out-of-equilibrium decays of the lightest right-handed neutrino generate a lepton asymmetry that is subsequently converted into a baryon asymmetry through sphaleron processes.

Flavor symmetries provide a well-motivated framework to explain neutrino masses and lepton mixing patterns \cite{Grossman:1998jj, Barman:2025bru, Chauhan:2023faf, Priya:2025khf, Priya:2026ytf, Tapender:2023kdk, Tapender:2025cfc}. But, conventional flavor models exhibit several drawbacks. The effective Lagrangian in such approaches typically involves a large number of flavon fields, leading to an extended parameter space. The vacuum alignment of these flavons plays a central role in determining the fermion mass matrices, thereby strongly constraining the model predictions and introducing sensitivity to alignment assumptions. Moreover, auxiliary symmetries are frequently required to suppress unwanted terms, which further increases the theoretical complexity. Most importantly, the flavor symmetry breaking sector introduces many undetermined parameters, significantly undermining the minimality and predictive power of the framework. By contrast, models based on modular symmetry provide a more economical and predictive alternative. In the modular symmetry framework, the introduction of flavon fields is not mandatory, and the breaking of flavor symmetry is achieved solely through the vacuum expectation value of the complex modulus $\tau$. Consequently, the number of free parameters is substantially reduced, resulting in a more constrained and predictive description of fermion masses and mixing \cite{Feruglio:2017spp, Ohki:2020bpo, deAdelhartToorop:2011re, Feruglio:2021dte, Kashav:2024lkr, Singh:2024imk, Kashav:2021zir, Kumar:2023moh,Mishra:2023ekx}. Furthermore, modular symmetry is motivated by string compactifications. The most distinctive aspect of this framework is that the modular symmetry acts not only on the fields, such as leptons and the Higgs field, but also on the coupling constants themselves, which transform non-trivially as a modular form under the modular group.

\begin{table*}[t]
\small
    \centering
    \caption{The neutrino oscillation data used in the numerical analysis taken from NuFIT 6.1. The central values of the charged lepton mass ratios are taken from~\cite{Esteban:2024eli, ParticleDataGroup:2024cfk,NuFit61}. During the scan of the model parameter space, the uncertainties in these ratios as fixed at 0.1\% of their respective central values~\cite{Xing:2007fb}.}
    \renewcommand{\arraystretch}{1} 
    \resizebox{\textwidth}{!}{\begin{tabular}{|l| c c c c|} 
        \hline
        Parameter & best-fit$\pm1\sigma$ range (NH) & best-fit$\pm1\sigma$ range (IH) & $3\sigma$ range (NH) & $3\sigma$ range (IH) \\
        \hline
        $\sin^2\theta_{12}$ & $0.308^{+0.0067}_{-0.0066}$ & $0.308^{+0.0067}_{-0.0066}$ & $0.289 - 0.329$ & $0.289 - 0.329$ \\
        \hline
        $\sin^2\theta_{23}$ & $0.470^{+0.017}_{-0.014}$ & $0.550^{+0.013}_{-0.016}$ & $0.435 - 0.584$ & $0.439 - 0.584$ \\
        \hline
        $\sin^2\theta_{13}$ & $0.02248^{+0.00055}_{-0.00059}$ & $0.02262^{+0.00057}_{-0.00056}$ & $0.02064 - 0.02418$ & $0.02093 - 0.02441$ \\
        \hline
        $\Delta m^2_{3l} \times 10^{-3} \text{eV}^2$ & $2.511^{+0.021}_{-0.020}$ & $-2.483^{+0.020}_{-0.020}$ & $2.450 - 2.576$ & $-2.547 - -2.421$ \\
        \hline
        $\Delta m^2_{21} \times 10^{-5} \text{eV}^2$ & $7.53^{+0.094}_{-0.10}$ & $7.53^{+0.094}_{-0.10}$ & $7.26 - 7.82$ & $7.23 - 7.82$ \\
        \hline
        $m_e/m_\mu$ & $0.004737$ & $0.004737$ & -- & -- \\
        \hline
        $m_\mu/m_\tau$ & $0.058823 $ & $0.058823$ & -- & -- \\
        \hline
    \end{tabular}}
    \label{tab:neutrino_data}
\end{table*}

The framework introduced by Qu and Ding extends this idea to non-holomorphic modular symmetries, where Yukawa couplings with negative modular weights are allowed through polyharmonic Maa{\ss} forms~\cite{Qu:2024rns}. Non-holomorphic modular symmetries have been studied by various authors in the context of different neutrino mass generation mechanisms, including the type-I~\cite{Nanda:2025lem}, type-II~\cite{Nomura:2024atp}, and type-III~\cite{Priya:2025wdm} seesaw mechanisms, and more~\cite{Dey:2025zld,Kumar:2024uxn,Kumar:2025nut,Ding:2024inn,Nomura:2024vzw,Li:2024svh,Loualidi:2025tgw,Li:2025kcr,Nomura:2024nwh,Okada:2025jjo,Abbas:2025nlv,Gao:2025jlw,Jangid:2025thp,Nomura:2025ovm,Nomura:2025raf,Zhang:2025dsa,Nasri:2026nbf,Tapender:2026ets,Zhang:2026kyy,Majhi:2026jdk}. In modular invariant flavor models, the fixed points of the modular symmetry are special values of the complex modulus $\tau$ that remain invariant under non-trivial modular transformations. When the modulus acquires a vacuum expectation value (VEV), the full modular symmetry is generally broken. However, at these fixed points, the symmetry breaking is incomplete and a subgroup of the original modular group remains unbroken. This surviving subgroup is referred to as the residual flavor symmetry. Each fixed point corresponds to a specific residual symmetry group, determined by the modular transformation that leaves $\tau$ invariant. In the case of level-3 modular symmetry based on the $A_4$ group, the fixed point $\tau = i$ preserves a $Z_2$ residual symmetry, while the fixed point $\tau = e^{2\pi i/3}$ preserves a $Z_3$ residual symmetry~\cite{Ding:2019gof, Okada:2020ukr}. These residual symmetries impose strong constraints on the allowed Yukawa couplings and mass matrices, leading to predictive structures for fermion masses and lepton mixing.

In this work, we investigate the type-III seesaw mechanism within the framework of non-holomorphic modular symmetry. We focus on the complex modulus $\tau$ in the vicinity of the modular fixed points, allowing for small deviations from the exact fixed-point values, which significantly reduces the number of free parameters in the framework. A comprehensive $\chi^{2}$ analysis is performed to examine the compatibility of these constrained configurations with current neutrino oscillation data. We find that only a restricted set of modular fixed points, together with small deviations around them, can successfully reproduce the observed neutrino oscillation data, while the remaining fixed points are incompatible with current experimental constraints. The corresponding numerical results are presented and discussed in detail.

The paper is organized as follows. In Section \ref{model_and_formalism}, the model is presented, and the procedure for diagonalizing the active neutrino mass matrix is discussed. The numerical analysis and the corresponding results are given in Section \ref{Numerical Analysis and Discussion}. Leptogenesis is addressed in Section \ref{section5}. Finally, the conclusions are presented in Section \ref{section6}.

\section{Modular symmetry and fixed Points} \label{fixed_point}
The modular symmetry can be interpreted as the origin of flavor symmetry. The modular transformation acts on the complex modulus $\tau$ in the upper half plane through linear fractional transformations
\begin{equation}
\tau \rightarrow \frac{a\tau+b}{c\tau+d}, \qquad a,b,c,d\in\mathbb{Z}, \quad ad-bc=1,
\end{equation}
which form the projective group $\mathrm{PSL}(2,\mathbb{Z})=\mathrm{SL}(2,\mathbb{Z})/\{\pm I\}$. 
The fundamental domain $\mathcal{F}$ of the modular group is defined by
\begin{equation}
\mathcal{F}=\left\{\tau\in\mathbb{C}\,\big|\,\mathrm{Im}(\tau)>0,\ |\tau|\ge 1,\ |\mathrm{Re}(\tau)|\le \tfrac12\right\}.
\end{equation}
The group is generated by two elements
\begin{equation}
S:\tau\to -\frac{1}{\tau}, \qquad T:\tau\to\tau+1,
\end{equation}
satisfying $S^2=(ST)^3=1$.

A modular form $Y_r^{(k)}(\tau)$ of weight $k$ is a function of modulus $\tau$, transforming in the representation $r$ of a finite modular group, that satisfies
\begin{equation}
Y_r^{(k)}(\gamma\tau)=(c\tau+d)^k\,\rho_r(\gamma)\,Y_r^{(k)}(\tau).
\label{eq:trans1}
\end{equation}
On the other hand, a field $\Phi$ with weight $k$ and representation $r$ also transforms as 
\begin{equation}
\Phi (\gamma \tau) = (c \tau + d)^{-k} \, \rho_r(\gamma) \, \Phi (\tau).
\end{equation}
Then a Lagrangian is invariant under modular symmetry if all the terms are invariant under these transformations.

\subsection{Fixed Points}

A point $\tau_0$ is called a \textit{fixed point} if it remains invariant under a nontrivial modular transformation $\gamma_0$,
\begin{equation}
\gamma_0\tau_0=\tau_0.
\end{equation}

Solving the fixed-point condition for a general transformation
\begin{equation}
\gamma_0=
\begin{pmatrix}
a_0 & b_0 \\
c_0 & d_0
\end{pmatrix},
\end{equation}
leads to a quadratic equation for $\tau_0$. Requiring $\tau_0\in\mathcal{F}$ restricts the trace of the transformation matrix to $|a_0+d_0|<2$, which gives only a few nontrivial solutions. As a result, the fundamental domain contains four inequivalent fixed points:

\begin{itemize}
\item \textbf{$S$-invariant point}
\begin{equation}
\tau_S=i,
\end{equation}
which is fixed under the transformation $S$.

\item \textbf{$ST$- and $TS$-invariant points}
\begin{equation}
\tau_{ST}=-\frac{1}{2}+i\frac{\sqrt{3}}{2}, 
\qquad
\tau_{TS}=\frac{1}{2}+i\frac{\sqrt{3}}{2},
\end{equation}
which are related by the action of $T$.

\item \textbf{$T$-invariant cusp}
\begin{equation}
\tau_T=i\infty.
\end{equation}
\end{itemize}

Thus, only these four points have nontrivial stabilizers inside the fundamental domain.

\subsection{Modular Forms at Fixed Points}

A modular form transforms as Eq.~\eqref{eq:trans1} under a modular symmetry.
At a fixed point $\tau_0$, this relation gives
\begin{equation}
\rho_r(\gamma_0)\,Y_r(\tau_0)=(c_0\tau_0+d_0)^{-k}Y_r(\tau_0),
\end{equation}
which shows that the modular forms at fixed points are eigenvectors of the representation matrix of the stabilizer transformation.

If the modulus $\tau$ takes one of these special values, the full modular symmetry is broken to a residual subgroup generated by the corresponding stabilizer ($S$, $ST$, or $T$). This residual symmetry constrains the structure of modular forms and is important in flavor model building. In our analysis, we consider several specific values of the modulus $\tau$. Among them, the point $\tau=i$ lies inside the fundamental domain, while the values
\[
\tau=\frac{1+i}{2}, \qquad 
\tau=\frac{3+i}{2}, \qquad 
\tau=1, \qquad 
\tau=-1
\]
do not belong to the fundamental domain but are related to special points through modular transformations. More precisely, these points are mapped into representatives inside the fundamental domain by elements of the modular group \(SL(2,\mathbb{Z})\). In particular, \(\tau=1\) and \(\tau=-1\), although lying on the real axis, are fixed points of non-trivial modular transformations, namely \(T^2S\) and \(ST^2\), respectively, and hence possess non-trivial stabilizers. The points \(\tau=\frac{1+i}{2}\) and \(\tau=\frac{3+i}{2}\) can be mapped into the fundamental domain by suitable combinations of the \(S\) and \(T\) generators. Since modular forms transform covariantly under these operations, the values of modular forms at these points are not independent but are determined by their values at the corresponding points inside the fundamental domain. Therefore, these \(\tau\) values are `special' in the sense that they are related by modular transformations to points with enhanced symmetry (such as fixed points or cusps of the modular group), and consequently inherit symmetry properties that constrain the structure of modular forms evaluated at those points. We include these values in our study to examine their effects on the structure of modular forms and the resulting phenomenological predictions.

\section{Model and Formalism} \label{model_and_formalism}
We extend the SM particle content by introducing a hyperchargeless fermion triplet, which implements the type-III seesaw mechanism. The charge assignments under modular group $A_4$ and modular weights($k_I$) for the model are shown in Table \ref{tab:charges}. The model content is same as in Ref~\cite{Priya:2025wdm}. 

The triplet fermions \(\Sigma_i\) with i = 1,2,3, can be presented in \(SU(2)\) basis as
\begin{equation}
    \Sigma_i = \begin{pmatrix}
        \frac{\Sigma_i^0}{\sqrt{2}} & \Sigma_i^+ \\
        \Sigma_i^- & -\frac{\Sigma_i^0}{\sqrt{2}}
    \end{pmatrix},
\end{equation}
where $\Sigma^0$ and $\Sigma^\pm$ corresponds to neutral and charged fermions respectively.

The modular invariant Lagrangian is given as
\begin{equation}
\begin{split}
    -\mathcal{L} =\; & \alpha\, \bar{L}\, H\, e_R\, Y_3^{(-2)} + \beta\, \bar{L}\,  H\, \mu_R\, Y_3^{(0)} 
    + \gamma\,\bar{L}\, H\,  \tau_R\, Y_3^{(0)} \\[5pt]
    & + g_1\, \bar{L}\, \Sigma_i\, H\,  Y_1^{(-2)} + g_2\, ( \bar{L}\,\Sigma_i )_{\text{sym}}\, H\,  Y_3^{(-2)} 
    + g_2'\, (\bar{L}\, \Sigma_i )_{\text{asym}}  \, H\,  Y_3^{(-2)}\\[5pt]
    & + M_0\, Tr[\Sigma_i\, \Sigma_i]\, Y_1^{(-4)} + M'\, Tr
    [\Sigma_i\, \Sigma_i]\, Y_3^{(-4)} + \text{H.C.}
    \label{eq:Lagrangian}
\end{split}
\end{equation}
where the subscript sym(asym) indicates that $A_4$ triplet is constructed symmetrically(anti-symmetrically) by triplets inside a bracket.

\noindent The corresponding charged lepton mass matrix is given as 
\begin{equation}
    M_L = \text{Diag}(\alpha,\beta,\gamma) 
    \begin{pmatrix}
        Y_{31}^{(-2)} & Y_{33}^{(-2)} & Y_{32}^{(-2)} \\
        Y_{32}^{(0)} & Y_{31}^{(0)} & Y_{33}^{(0)} \\
        Y_{33}^{(0)} & Y_{32}^{(0)} & Y_{31}^{(0)}
    \end{pmatrix}v,
\end{equation}

\noindent here $v$ is the $vev$ of the Higgs field. The charged lepton mass matrix $ M_L$ contains three real parameters $ \alpha, \beta$ and $\gamma $, which can be suitably chosen to reproduce the observed charged lepton masses. The diagonalization of $M_L$ is given as $U_L^\dagger M_L^\dagger M_L U_L = \text{Diag}{(|m_e|^2, |m_\mu|^2, |m_\tau|^2)}$.  

The second lines of the Lagrangian Eq.~\eqref{eq:Lagrangian} induce a Dirac mass term between the SM neutrino and extra neutral fermion $\Sigma^0$.
The corresponding Dirac mass matrix is given as 
\begin{equation}\label{mdyukawa}
    M_D =  v \begin{pmatrix}
        g_1 Y_1^{(-2)} + 2 \frac{g_2}{3} Y_{31}^{(-2)} &   (\frac{g_2'}{2}-\frac{g_2}{3}) Y_{33}^{(-2)}&  -(\frac{g_2'}{2}+\frac{g_2}{3}) Y_{32}^{(-2)} \\
         -(\frac{g_2'}{2}+\frac{g_2}{3})Y_{33}^{(-2)} & 2 \frac{g_2}{3} Y_{32}^{(-2)} & g_1 Y_1^{(-2)}+(\frac{g_2'}{2}-\frac{g_2}{3}) Y_{31}^{(-2)} \\
         (\frac{g_2'}{2}-\frac{g_2}{3})Y_{32}^{(-2)} & g_1 Y_1^{(-2)}- (\frac{g_2'}{2}+\frac{g_2}{3}) Y_{31}^{(-2)} & 2 \frac{g_2}{3} Y_{33}^{(-2)}
    \end{pmatrix}.
\end{equation}

\begin{table}[t]
\centering
\caption{Field content and charge assignments of Type-III seesaw mechanism under $A_4$ group and modular weights.}
\begin{tabular}{|c|cccccc|}
\hline
 & $\bar{L}$ & $e_R$&$\mu_R$&$\tau_R$ & $\Sigma_i$ & $H$   \\
\hline
$SU(2)$ & $2$ & $1$ &$1$ &$1$ & $3$ & $2$  \\
\hline
$A_4$ & $3$ & $1$ & $1''$& $1'$ & $3$ & $1$   \\
\hline
$k_I$ & $0$ & $-2$&0&$0$ & $-2$ & $0$   \\
\hline
\end{tabular}
\label{tab:charges}
\end{table}

The invariant modular Lagrangian for the Majorana term is given as 
\begin{equation}
    -L_\Sigma = M_0  Tr[\Sigma_i \Sigma_i] Y_1^{(-4)} + M'  Tr[\Sigma_i \Sigma_i] Y_3^{(-4)} + \text{H.C.}
\label{Ln}
\end{equation}
\noindent The right-handed Majorana neutrino mass matrix is given as

\begin{equation}
    M_\Sigma =  \begin{pmatrix}
       M_0 Y_1^{(-4)}+ 2 M' Y_{31}^{(-4)} &   -M'Y_{33}^{(-4)}& - M' Y_{32}^{(-4)} \\
         - M' Y_{33}^{(-4)} & 2M' Y_{32}^{(-4)} &  M_0 Y_1^{(-4)}-M'Y_{31}^{(-4)} \\
         -M'Y_{32}^{(-4)} &  M_0 Y_1^{(-4)}-M'Y_{31}^{(-4)} & 2 M'Y_{33}^{(-4)}
    \end{pmatrix}.
\end{equation}
The mass matrix $M_\Sigma$, can be diagonalized by using the unitary matrix, such that, $U_R$ as $U_R^TM_\Sigma U_R = \text{diag}(M_{\Sigma 1},M_{\Sigma 2},M_{\Sigma 3})$.
\noindent Finally, the active neutrino mass matrix is obtained by the Type-III seesaw as follows
\begin{equation}
M_\nu = -M_D M_\Sigma^{-1} M_D^T.
\label{neutrinomassmatrix}
\end{equation}
The neutrino mass matrix given by Eqn. (\ref{neutrinomassmatrix}) is diagonalized using the relation $U_\nu^T M_\nu U_\nu = \text{diag}(m_{\nu_1}, m_{\nu_2}, m_{\nu_3})$. The mixing matrix $U_{\rm PMNS}$ = $U_L^\dagger U_\nu$, since the charged lepton mass matrix is not diagonal in the flavor basis. Now, the mixing angle can be extracted from $U_{\rm PMNS}$ as 

\begin{equation}
    \sin^2 \theta_{13} = |U_{13}|^2, \quad
    \sin^2 \theta_{12} = \frac{|U_{12}|^2}{1 - |U_{13}|^2}, \quad {\rm and} \quad
    \sin^2 \theta_{23} = \frac{|U_{23}|^2}{1 - |U_{13}|^2}. 
\end{equation}
The Dirac $CP$-violating phase($\delta_{CP}$) can be determined from the PMNS matrix elements through the Jarlskog invariant, defined as
\begin{equation}
    J_{CP} = \text{Im}\left[U_{11} U_{22} U^{*}_{12} U^{*}_{21}\right] = s_{23} c_{23} s_{12} c_{12} s_{13} c_{13}^2 \sin\delta_{CP},
\end{equation}
where \( s_{ij} = \sin\theta_{ij} \) and \( c_{ij} = \cos\theta_{ij} \). In addition to $\delta_{CP}$, the Majorana $CP$ phases can be investigated using the PMNS matrix elements as
\begin{equation}
    I_1 = \text{Im}\left[U_{11}^* U_{12}\right] = c_{12} s_{12} c_{13}^2 \sin\left(\frac{\alpha_{21}}{2}\right),
\end{equation}
\begin{equation}
    I_2 = \text{Im}\left[U_{11}^* U_{13}\right] = c_{12} s_{13} c_{13} \sin\left(\frac{\alpha_{31}}{2} - \delta_{CP}\right).
\end{equation}

\section{Numerical Analysis and Discussion}
\label{Numerical Analysis and Discussion}

In this section, we carry out numerical analysis for the Type-III seesaw mechanism within the framework of non holomorphic modular symmetry, considering its behavior at the fixed points. To examine the consistency of our model with current experimental data, we have performed a $\chi^2$ analysis. The chi-square function used in analysis is shown as

\begin{equation}
\chi^2 = \sum_{i = 1}^7 \left( \frac{P_i - P_i^0}{\sigma_i} \right)^2, \label{chisq}
\end{equation}

\noindent here, $P_i$ denotes the values predicted by the model, while $P_i^{0}$ represents the best-fit values of the neutrino observables obtained from the NUFit 6.1 analysis summarized in Table~\ref{tab:neutrino_data}. The $\sigma_i$ corresponds to the $1\sigma$ experimental uncertainty associated with each observable. In this work, we consider seven observables: the three leptonic mixing angles ($\sin^2\theta_{13},~ \sin^2\theta_{12},~ \sin^2\theta_{23}$), the two neutrino mass-squared differences ($\Delta m_{21}^2$ and $\Delta m_{31}^2$), and two charged-lepton mass ratios ($m_e/m_\mu$ and $m_\mu/m_\tau$). The Dirac CP phase ($\delta_{CP}$) is not included as an input parameter due to its comparatively weak experimental constraints. The parameter space has been explored using a Markov Chain Monte Carlo (MCMC) technique, and a total of $10^{9}$ sample points have been generated using a Python-based implementation. In our analysis, the independent free parameters of the model are $\alpha/\beta$, $\gamma/\beta$, $g_1$, $g_2$, $g_{11}$, $M_0$, $m_1$, as the small expansion parameter $\epsilon$ and the phase $\phi$, over which we perform a systematic scan.

\begin{table*}[t]
    \centering
    \caption{The best-fit values for the neutrino oscillation parameters obtained from the $\chi^2$ analysis for NH.}
    \resizebox{\textwidth}{!}{\begin{tabular}{|c|cccccccccccc|}
        \hline
        & $\chi_{\rm min}^2$& $\sin^2\theta_{12}$ & $\sin^2\theta_{23}$ & $\sin^2\theta_{13}$ & $\delta_{CP}(^\circ)$ & $\Delta m^2_{21} (\text{eV}^2)$ & $\Delta m^2_{31} (\text{eV}^2)$ & $\Sigma_i \text{(eV)}$& ${\rm Re}[\tau]$ & ${\rm Im}[\tau]$& $\epsilon$ & $\phi$\\ 
        \hline
        $\frac{3+i}{2}$& 0.17 &0.307 & 0.466 & 0.022 & $126.73^o$ & $7.6 \times 10^{-5}$ & $2.51 \times 10^{-3}$& $0.099$ & $1.389$ & $0.533$ & $0.073$ & $0.80\pi$ \\
        \hline
        1&0.14&0.307 & 0.473 & 0.022 & $108.20^\circ$ & $7.5 \times 10^{-5}$ & $2.50 \times 10^{-3}$ & $0.112$ & $0.958$ & $0.034$ & $0.0542$ & $0.77\pi$ \\
        \hline
        -1 & 0.20 &0.308 & 0.474 & 0.022 & $252.22^\circ$ & $7.5 \times 10^{-5}$ & $2.50 \times 10^{-3}$ & $0.112$ & -$0.958$ & $ 0.035$& $0.0540$& $-0.77\pi$\\
        \hline
    \end{tabular}}
    \label{tab:neutrino_params}
\end{table*}

\begin{table*}[t]
    \centering
    \caption{The best-fit values for the neutrino oscillation parameters obtained from the $\chi^2$ analysis for IH.}
    \resizebox{\textwidth}{!}{\begin{tabular}{|c|cccccccccccc|}
        \hline
        & $\chi_{\rm min}^2$&$\sin^2\theta_{12}$ & $\sin^2\theta_{23}$ & $\sin^2\theta_{13}$ & $\delta_{CP} (^\circ)$ & $\Delta m^2_{21} (\text{eV}^2)$ & $\Delta m^2_{31} (\text{eV}^2)$ & $\Sigma_i \text{(eV)}$& $\rm Re[\tau]$ & $\rm Im[\tau]$& $\epsilon$ & $\phi$\\
        \hline
        $\frac{3+i}{2} $& 1.08 & 0.310 & 0.545 & 0.022 & $299.91^o$ & $7.5 \times 10^{-5}$ & $-2.48 \times 10^{-3}$& $0.139$ & $1.380$ & $0.492$ &  $0.040$ & $-0.92\pi$   \\
        \hline
       1& 0.44 &  0.306 & 0.548 & 0.022 & $313.63^\circ$ & $7.5 \times 10^{-5}$ & $-2.48 \times 10^{-3}$ & $0.140$ &  0.923 & 0.022   & 0.079 &  $0.91\pi$ \\
       \hline
         -1 & 0.08 & 0.308 & 0.549 & 0.022 & $35.67^\circ$ & $7.5 \times 10^{-5}$ & $-2.48 \times 10^{-3}$ & $0.172$ & $-0.924$  & $0.019 $ &$ 0.078$ &$0.92\pi$\\
        \hline
    \end{tabular}}
    \label{tab:neutrino_params_ih}
\end{table*}

We have analyzed the model in the vicinity of the fixed points. However, only a subset of these fixed points is consistent with the current neutrino oscillation data. We present the correlation plots corresponding to the fixed points that are compatible with the experimental observations. The correlations are shown for the parameter space satisfying $\chi^2 < 10$, where the star ($\star$) denotes the best-fit value. For the fixed point $\tau_{\rm fixed}$, we performed a detailed scan of the nearby parameter space as $\tau = \tau_{\rm fixed}(1+\epsilon e^{i\phi})$ by varying the phase $\phi$ in the range $(-\pi,\pi)$ and the parameter $\epsilon$ in the range $(0,0.1)$. The model predictions were then tested against the existing neutrino oscillation data.
In the following, we summarize our results for each fixed point that can accommodate the observed data.

\subsection{At nearby $\dfrac{3+i}{2}$}
\begin{figure*}[t]
\centering
\includegraphics[width=0.9\linewidth]{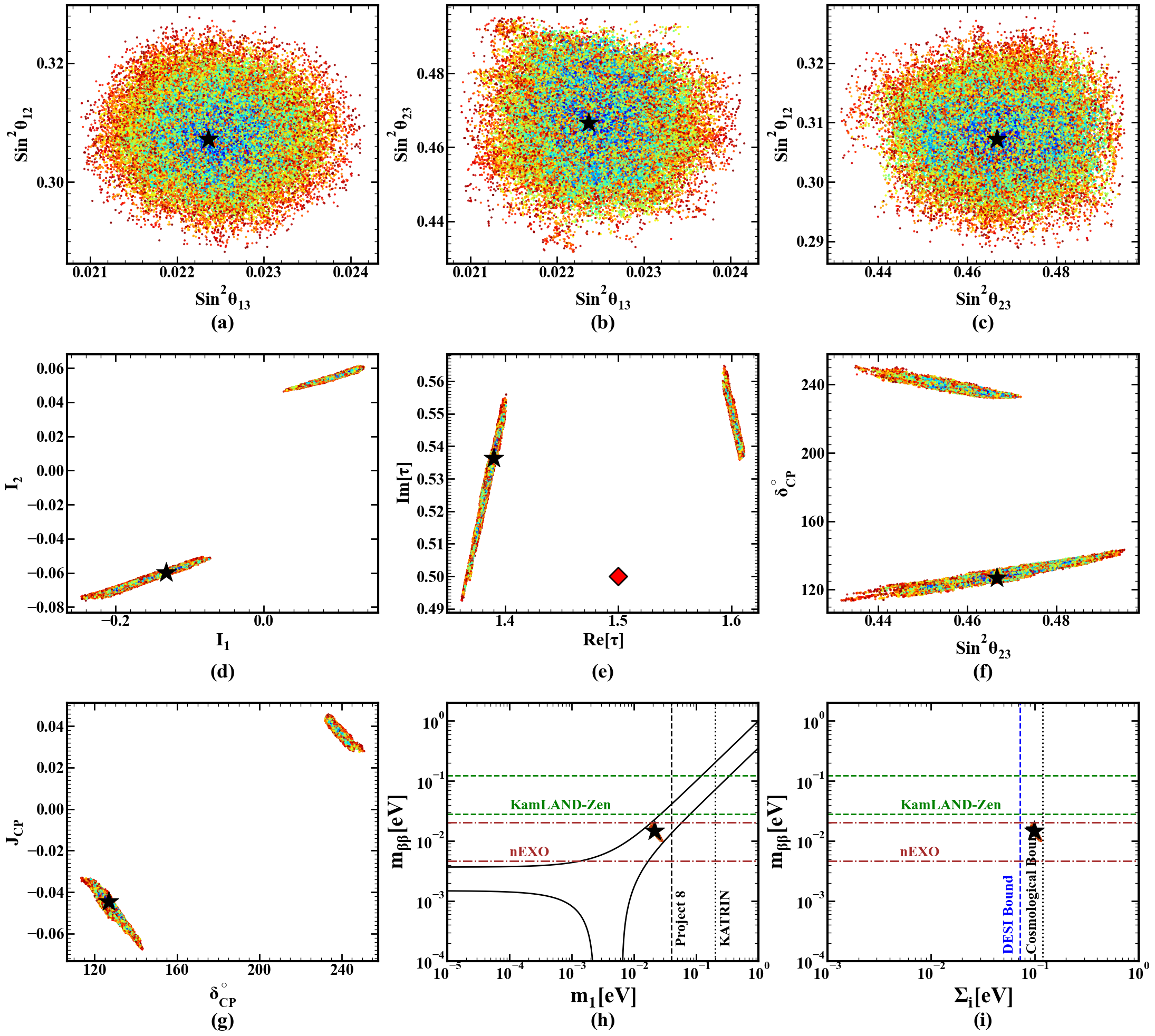}
\includegraphics[width=0.4\linewidth]{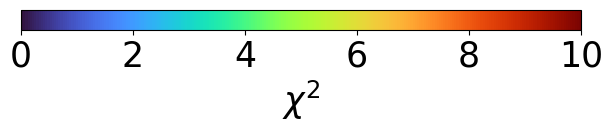}
\caption{Correlation between mixing angles, real and imaginary part of $\tau$, Jarlskog invaraint , Dirac-type CP phase, CP invariants are shown. Additionally effective majorana mass with respect to lightest neutrino mass and sum of neutrino mass is also presented with minimum chi square at 0.072 for nearby fixed point $\tau = \frac{3+i}{2}$ for NH}
\label{Fig1_3}
\end{figure*}

We find a minimum $\chi^2$ value of $0.17$ for the nearby region of fixed point $\frac{3+i}{2}$, and the corresponding best-fit values are listed in Table~\ref{tab:neutrino_params}. Fig.~\ref{Fig1_3}(a), ~\ref{Fig1_3}(b) and \ref{Fig1_3}(c) illustrate the quality of the fit to the neutrino oscillation observables. All the three mixing angles are compatible with current neutrino oscillation data at $1\sigma$ level. The correlation between the CP invariants $I_1$ and $I_2$ is shown in Fig.~\ref{Fig1_3}(d). Fig.~\ref{Fig1_3}(e) shows the correlation between the real and imaginary parts of the complex modulus $\tau$, illustrating the deviation from the selected fixed point. The best-fit value is located at $\rm Re[\tau] = 1.389$ and $\rm Im[\tau] = 0.533$, and is marked by an star ($\star$) in the figure. The exact fixed point is indicated in the figure by a red diamond. The correlation between the Dirac CP phase $\delta_{CP}$ and the atmospheric mixing angle is shown in Fig.~\ref{Fig1_3}(f), from which we observe that the model predicts the best-fit value of $\sin^2\theta_{23}$ in the first octant. This prediction can be tested in future long baseline neutrino oscillation experiments(LBL)~\cite{2852068,DUNE:2020fgq,Hyper-Kamiokande:2018ofw}. The predicted range of the Dirac CP phase is consistent with the current global-fit results. Fig.~\ref{Fig1_3}(g) shows the correlation between Jarlskog invariant and $\delta_{CP}$. The correlation between the effective Majorana mass $m_{\beta\beta}$ and the lightest neutrino mass is presented in Fig.~\ref{Fig1_3}(h). The predicted values of $m_{\beta\beta}$ are compatible with the KamLand~\cite{KamLAND-Zen:2024eml} and current nEXO sensitivity~\cite{nEXO:2021ujk}, and the lightest neutrino mass($m_1$) lies within the detectable range of KaTRIN($<0.45eV$ at 90\% Confidence level(CL))~\cite{KATRIN:2022ayy} and the Project~8 experiment~\cite{Project8:2022wqh}. Fig.~\ref{Fig1_3}(i) shows the correlation between $m_{\beta\beta}$ and the sum of neutrino masses($\Sigma_i$). The predicted sum of neutrino masses is consistent with the current cosmological bound$(<0.12eV$ at 95\% CL)~\cite{Planck:2018vyg}. The best-fit value of the sum of neutrino masses is $\Sigma_i = 0.099\,\text{eV}$, corresponding to a minimum chi-square value $\chi^2_{\min} = 0.17$. The model predicts an allowed range for the sum of neutrino masses of $(0.092\text{-}0.119)\,\text{eV}$. For inverted hierarchy, neutrino observables are compatible with the current experimental data with $\chi^2_{min} = 1.08$. The sum of neutrino masses exceeds the cosmological bound as model predicts the $\Sigma_i = 0.139eV$.

\subsection{At nearby fixed point $1$}
\begin{figure*}[t]
\centering
\includegraphics[width=0.9\linewidth]{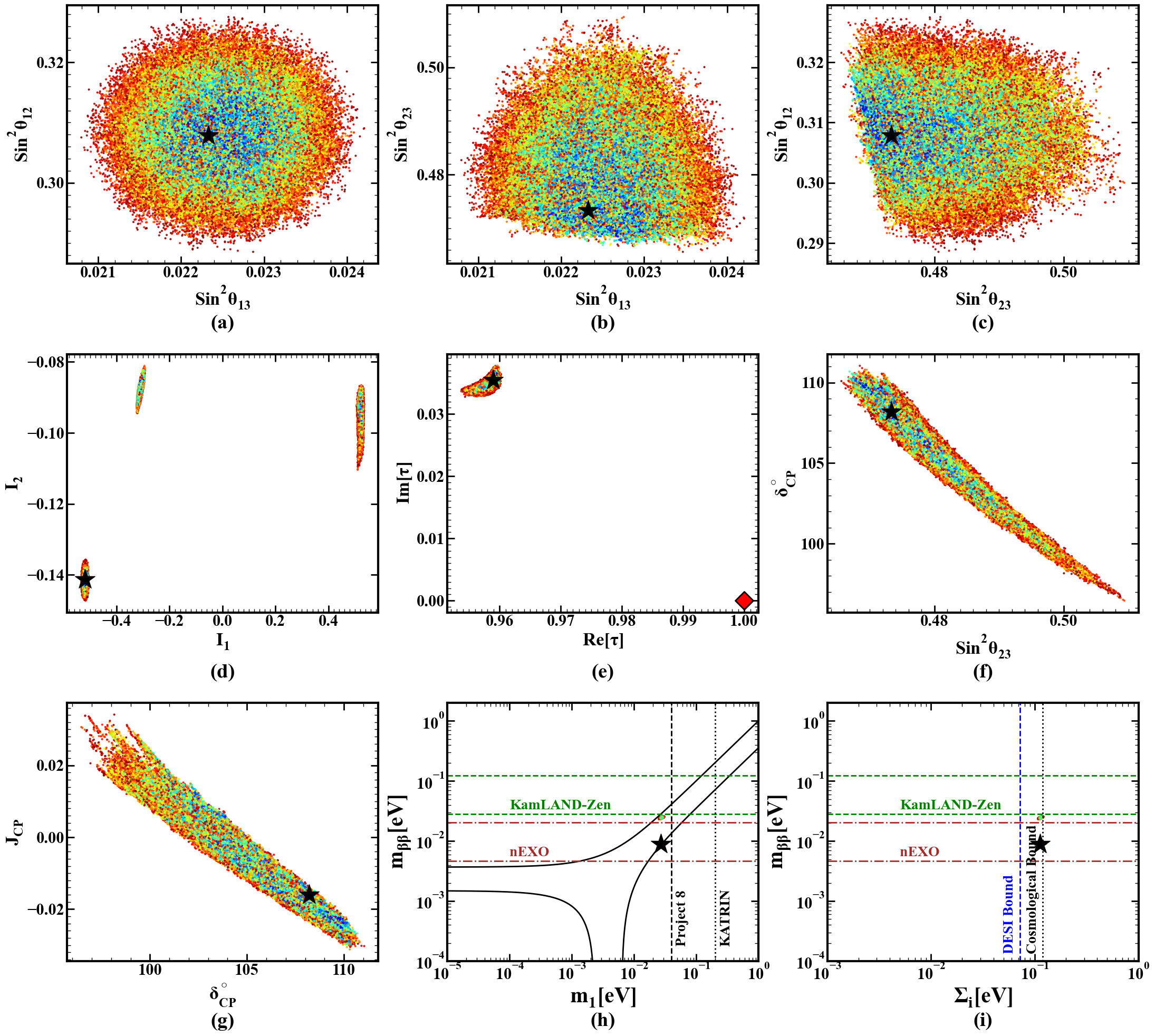}
\includegraphics[width=0.4\linewidth]{color_bar.png}
\caption{Correlation between mixing angles, real and imaginary part of $\tau$, Jarlskog invariant, Dirac-type CP phase, CP invariants are shown. Additionally effective majorana mass with respect to lightest neutrino mass and sum of neutrino mass is also presented with minimum chi square at 0.14 for nearby fixed point $\tau = 1$}
\label{Fig_nearby_7}
\end{figure*}

We also obtain a chi-square minimum in the vicinity of the fixed point $\tau = 1$. In this region, the minimum value of $\chi^2$ is $0.14$, and the corresponding best-fit values are listed in Table~\ref{tab:neutrino_params}. Fig.~\ref{Fig_nearby_7}(a), ~\ref{Fig_nearby_7}(b) and ~\ref{Fig_nearby_7}(c) shows that all three leptonic mixing angles predicted by the model are compatible with the current experimental data at 1 $\sigma$ level. In particular, the atmospheric mixing angle $\theta_{23}$ is clearly predicted to lie in the first octant. These predictions can be tested in future LBL neutrino experiments\cite{2852068,DUNE:2020fgq,Hyper-Kamiokande:2018ofw}. The correlations among the CP invariants($I_1~\text{and}~ I_2$) are displayed in Fig.~\ref{Fig_nearby_7}(d). Fig.~\ref{Fig_nearby_7}(e) illustrates the deviation of the real and imaginary parts of the complex modulus $\tau$ from the fixed point $\tau = 1$ with best fit at $\rm Re[\tau] = 0.958$ and $\rm Im[\tau] = 0.034$ corresponding to $\chi^2_{min} = 0.14$, and is marked by an star ($\star$) in the figure. The exact fixed point is indicated in the figure by a red diamond. The correlation between the Dirac CP phase $\delta_{\rm CP}$ and $\sin^2\theta_{23}$ is shown in Fig.~\ref{Fig_nearby_7}(f), which indicates that the model predicts CP violation, with $\delta_{\rm CP}$ lying predominantly in the second quadrant. The correlation between the Jarlskog invariant $J_{\rm CP}$ and $\delta_{\rm CP}$ is also presented in Fig.~\ref{Fig_nearby_7}(g). The correlation involving the $m_{\beta\beta}$ with $m_1$, is shown in Fig.~\ref{Fig_nearby_7}(h). The predicted values of $m_{\beta\beta}$ are compatible with the sensitivity of the nEXO experiment~\cite{nEXO:2021ujk}, while the $m_{1}$ lies within the projected sensitivity of the Project~8 experiment~\cite{Project8:2022wqh}. The correlation between $m_{\beta\beta}$ and the sum of neutrino masses($\Sigma_i $) is also shown in Fig.~\ref{Fig_nearby_7}(i). The best-fit value of the sum of neutrino masses obtained in our analysis is compatible with the current cosmological upper bound~\cite{Planck:2018vyg}. We obtain a best-fit value of the sum of neutrino masses $\Sigma_i = 0.112\,\text{eV}$, corresponding to $\chi^2_{\min} = 0.14$. Allowing a wider parameter space defined by $\chi^2 < 10$, the predicted range for the sum of neutrino masses is $(0.105\text{-}0.133)\,\text{eV}$. The IH is not allowed in this model for such selection of fixed points at predict the $\chi^2_{min} = 0.44$. The sum of neutrino masses is $0.140eV$ corresponding to $\chi^2_{min}$, which shows incompatibility with cosmological bound provided by Planck's data. The relevant neutrino parameters for IH are summarized in Table.~\ref{tab:neutrino_params_ih}

\subsection{At nearby fixed point $-1$}
\begin{figure*}[t]
\centering
\includegraphics[width=0.9\linewidth]{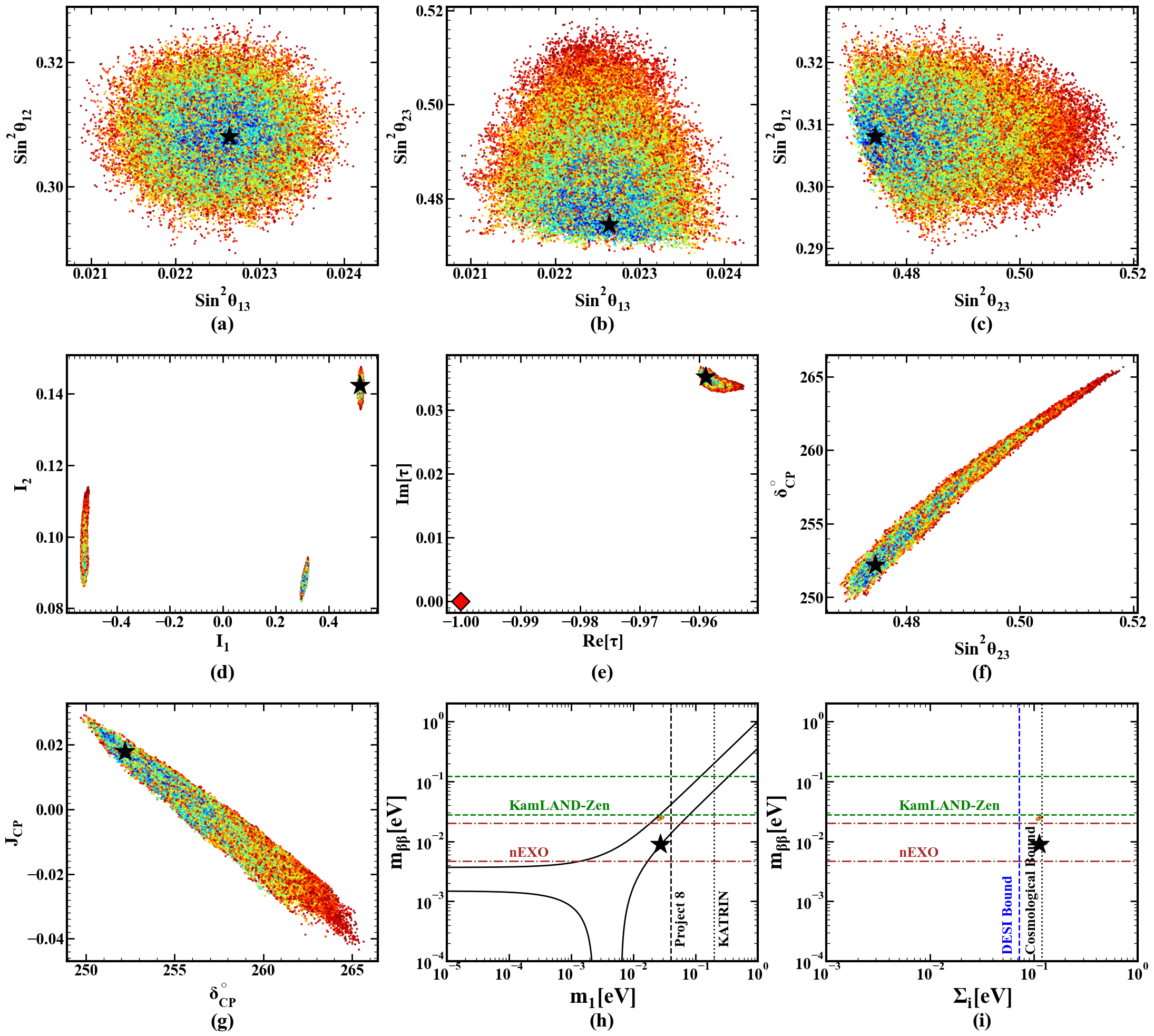}
\includegraphics[width=0.4\linewidth]{color_bar.png}
\caption{Correlation between mixing angles, real and imaginary part of $\tau$, Jarlskog invariant, Dirac-type CP phase, CP invariants are shown. Additionally effective majorana mass with respect to lightest neutrino mass and sum of neutrino mass is also presented with minimum chi square at 0.10 for nearby fixed point $\tau = -1$}
\label{Fig_nearby_8}
\end{figure*}

We also obtain a chi-square minimum in the vicinity of the fixed point $\tau = -1 $. The minimum value of the chi-square function is $\chi^2_{\rm min} = 0.20$, and the corresponding best-fit values are listed in Table~\ref{tab:neutrino_params}. Fig.~\ref{Fig_nearby_8}(a),~\ref{Fig_nearby_8}(b) and ~\ref{Fig_nearby_8}(c) shows that the model predicts all three leptonic mixing angles in good agreement with the current experimental data. In particular, the atmospheric mixing angle $\sin^2\theta_{23}$ is predicted to lie in the first octant, which can be tested in future LBL neutrino oscillation experiments~\cite{2852068,DUNE:2020fgq,Hyper-Kamiokande:2018ofw}. The correlations among the CP invariants are presented in Fig.~\ref{Fig_nearby_8}(d).Fig.~\ref{Fig_nearby_8}(e) illustrates the deviation of the real and imaginary parts of the complex modulus $\tau$ from the fixed point. The best–fit point is obtained at $\rm Re[\tau] = -0.958$ and $\rm Im[\tau] = 0.035$, corresponding to $\chi^2_{\min} = 0.20$. The exact fixed point is indicated in the figure by a red diamond. The correlation between the Dirac CP phase $\delta_{\rm CP}$ and $\sin^2\theta_{23}$ is displayed in Fig.~\ref{Fig_nearby_8}(f), from which one can clearly see that the model predicts strong CP violation together with a preference for the first octant of $\theta_{23}$. The correlation between the Jarlskog invariant $J_{\rm CP}$ and $\delta_{\rm CP}$ is shown in Fig.~\ref{Fig_nearby_8}(g). The $m_{\beta\beta}$ is plotted as a function of the lightest neutrino mass in Fig.~\ref{Fig_nearby_8}(h). The predicted values of $m_{\beta\beta}$ are consistent with the current nEXO bound~\cite{nEXO:2021ujk}, while $m_{1}$ lies within the projected sensitivity of the Project~8 experiment~\cite{Project8:2022wqh}. The correlation between $m_{\beta\beta}$ and the sum of neutrino masses $\Sigma_i$ is also shown in Fig.~\ref{Fig_nearby_8}(i). These predictions are consistent with the current cosmological bound~\cite{Planck:2018vyg}, with a best-fit value of the sum of neutrino masses $\Sigma_i = 0.112\,\text{eV}$ corresponding to a minimum chi-square value $\chi^2_{\min} = 0.20$. However, tension with the DESI bound($<0.072eV$ at 95\% CL) is observed~\cite{DESI:2024mwx}. The predicted range of the sum of neutrino masses is $(0.106\text{-}0.160)\,\text{eV}$. The IH is not compatible with the choice of this model and this nearby fixed point as it predict $\chi_{min}^2 = 0.089$. The sum of neutrino mass predicted by the model is $\Sigma_i = 0.172eV$, which is not compatible with the cosmological bound. The relevant neutrino parameters for IH are summarized in Table~\ref{tab:neutrino_params_ih}. Although the concept of modular fixed points is well established, their role in constraining leptogenesis within the present framework is non-trivial. In particular, they provide predictive benchmark regions where the heavy fermion spectrum and the CP structure are tightly correlated. While the neutrino oscillation observables are consistent with current experimental bounds, their significance in this framework extends beyond a purely low-energy description. At the fixed points, the Yukawa couplings entering the Type-III seesaw mechanism are strongly constrained by modular symmetry and neutrino data. These same couplings also govern the decay properties and CP asymmetries of the heavy fermion triplets, thereby establishing a direct connection between low-energy neutrino phenomenology and the dynamics of leptogenesis in the early Universe.

\section{Leptogenesis}\label{section5}
\noindent The observed baryon asymmetry of the Universe can be explained by the mechanism of leptogenesis, which operates in seesaw models through the decays of heavy right-handed neutrinos. When the temperature of the universe falls below the mass of the lightest heavy state, its decay occurs out of thermal equilibrium and generates a lepton asymmetry through CP-violating interactions. This lepton asymmetry is subsequently reprocessed into a baryon asymmetry by non-perturbative electroweak sphaleron transitions that violate baryon and lepton numbers while conserving $B-L$. The generation of such an asymmetry requires the fulfillment of the three Sakharov conditions: baryon number violation, C and CP violation, and departure from thermal equilibrium. In the Standard Model, baryon number violation arises at high temperatures due to sphaleron processes, while C and CP violation originate from weak interactions and complex Yukawa couplings. However, the electroweak phase transition within the Standard Model is a crossover rather than a strong first-order transition, and therefore does not provide sufficient departure from thermal equilibrium. As a result, the Standard Model alone cannot account for the observed baryon asymmetry of the Universe.

\noindent In modular symmetric frameworks, CP violation has a geometric origin and arises from the imaginary part of the complex modulus $\tau$. At specific modular fixed points, both the real and imaginary parts of $\tau$ are fixed, which uniquely determines the structure of the Yukawa couplings. In this work, the required CP asymmetry is generated through the out-of-equilibrium decays of the lightest fermion triplet, whose decay properties are fully governed by the modular fixed-point values of $\tau$. The CP asymmetry generated by the decay of lightest fermion triplet is given as 
\begin{equation}
    \epsilon_i = -\sum_{j = 2,3} \frac{3}{2}\frac{M_{\Sigma_i}}{M_{\Sigma_j}} \frac{\Gamma_j}{M_{\Sigma_j}} I_j \frac{V_j - 2 S_j}{3},
    \end{equation}
\noindent where, 
\begin{equation}
    I_j = \frac{Im[(\tilde{Y}^\dagger \tilde{Y})_{ij}^2]}{(\tilde{Y}^\dagger \tilde{Y})_{ii}(\tilde{Y}^\dagger \tilde{Y})_{jj}},
\end{equation}
where, $\tilde{Y} = Y U_L U_R$ and $Y = \frac{M_D}{v}$ and $V_j$ and $S_j$ are the loop factors associated with the vertex and self-energy corrections, respectively, given by 

\begin{equation}
    V_j = \frac{M_{\Sigma_j}^2 (M_{\Sigma_j}^2 - M_{\Sigma_i}^2)}
{(M_{\Sigma_j}^2 - M_{\Sigma_i}^2)^2 + M_{\Sigma_i}^2 \Gamma_{\Sigma_j}^2},
\end{equation}
\quad
\begin{equation}
    S_j = 2 \frac{M_{\Sigma_j}^2} {M_{\Sigma_i}^2} \left( \left( 1 + \frac{M_{\Sigma_j}^2}{M_{\Sigma_i}^2} \right)  
\ln \left( 1 + \frac{M_{\Sigma_j}^2}{M_{\Sigma_i}^2} \right) - 1 \right).
\end{equation}
In the hierarchical limits($M_{\Sigma _1}< M_{\Sigma_2} < M_{\Sigma _3} $), the loop factors reduce to unity \cite{Hambye:2012fh, Albright:2003xb,Mishra:2022egy}. Further, $\Gamma_j$ represents the decay width of the triplet fermion and can be expressed as 

\begin{equation}
\Gamma_{\Sigma_j} = \left( \frac{|(\tilde{Y}^\dagger \tilde{Y})_{jj}|}{8\pi} \right) M_{\Sigma_j}.
\end{equation}
\noindent The Boltzmann equation (BEs) plays an important role in tracking the evolution of lepton asymmetry as the universe gradually cools down over time. The relevant coupled BEs are given by
\begin{equation}
s \textbf{H} z \frac{dY_\Sigma}{dz} = -\gamma_D \left( \frac{Y_\Sigma}{Y^{\text{eq}}_\Sigma} - 1 \right) - 2 \gamma_A \left( \frac{Y_\Sigma^2}{(Y^{\text{eq}}_\Sigma)^2} - 1 \right),
\label{ysigma}
\end{equation}

\begin{equation}
s \textbf{H} z \frac{dY_{B-L}}{dz} = -\gamma_D \, \epsilon_\Sigma \left( \frac{Y_\Sigma}{Y^{\text{eq}}_\Sigma} - 1 \right) - \frac{Y_{B-L}}{Y^{\text{eq}}_l} \left( \frac{\gamma_D}{2} + \gamma^{\text{sub}}_\Sigma \right),
\end{equation}
where, $Y_{\Sigma a}$ = $\frac{n_a(z)}{s(z)}$ represents the number density of particle species $a$. $s(z)$ is the entropy density given as $s(z)= 0.44 g_* T^3$, $z$ is a dimensionless parameter varying inversely with the temperature of the universe, given as $z \equiv \frac{M_{\Sigma_i}}{T}$. $\textbf{H}$ represents the Hubble expansion rate, given as $\textbf{H} = 1.66 \, \frac{\sqrt{g_*} \, T^2}{M_{\text{Pl}}}$. $\gamma$ denotes the reaction density of processes under consideration, and `D' denotes the decay processes and given as 
\begin{equation}
    \gamma _D (z) = s(z) Y_\Sigma^{eq} \Gamma _{\Sigma} \frac{K_1(z)}{K_2(z)},
\end{equation}

\noindent where $K_1(z)$ and $K_2(z)$ are the modified Bessel functions and $\gamma _A$ in Eqn. (\ref{ysigma}) denotes the gauge annihilation processes and represented as 
\begin{equation}
\gamma _A(z) = \frac{M_{\Sigma_1} T^3}{32\pi^3} e^{-2z}  
\left[
\frac{111 g^4}{8\pi} + \frac{3}{2z} \left( \frac{111 g^4}{8\pi} + \frac{51 g^4}{16\pi} \right) + \mathcal{O}\left(\frac{1}{z^2}\right)
\right],
\end{equation}

\noindent where, $g$ is the typical gauge coupling. The $\gamma^{\text{sub}}$ is for washout effects by $\Delta$L = 2 processes. The equilibrium yields are given by

\begin{equation}
    Y_\Sigma^{eq} = \frac{135 g_\Sigma}{16 \pi^4 g_*} z^2 K_2(z), \quad Y_l^{eq} = \frac{3}{4} \frac{45 \zeta(3) g_l}{2 \pi^4 g_*},
\end{equation}

\noindent where $g_l = 2$, $g_\Sigma = 2$ and $g_* = 106.75$. For a lepton asymmetry to survive, the decays of heavy right-handed neutrinos must occur out of thermal equilibrium. This condition is satisfied when the decay rate of the heavy fermions is not significantly larger than the Hubble expansion rate of the Universe at a temperature equal to their mass, $T = M_{\Sigma_i}$. Under such circumstances, the generated lepton asymmetry is not completely washed out. The departure from thermal equilibrium is commonly quantified by the decay parameter $k$, defined as the ratio of the decay width of the heavy fermion to the Hubble expansion rate evaluated at $T = M_{\Sigma_1}$,
\[
k = \frac{\Gamma_\Sigma}{\mathbf{H}(T = M_{\Sigma_1})}
    = \frac{\tilde{m}_i}{m^*}.
\]
Here, the effective neutrino mass $\tilde{m}_i$ is given by
\[
\tilde{m}_i = \frac{(\tilde{Y}^\dagger \tilde{Y})_{ii} v^2}{M_{\Sigma_i}},
\]
While $m^*$ denotes the equilibrium neutrino mass scale evaluated at $T = M_{\Sigma_1}$, the washout parameter $k$ allows one to distinguish three different regimes: $k \ll 1$ corresponds to the weak washout regime, $k \sim 1$ to the intermediate washout regime, and $k \gg 1$ to the strong washout regime~\cite{Davidson:2008bu}. Using the best-fit values of the model parameters, we find that the model consistently lies in the strong washout regime. In this regime, the washout effects are dominated by inverse decay processes, while the contributions from $\Delta L = 1$ and $\Delta L = 2$ scattering processes are subdominant and can be safely neglected. Therefore, following Refs.~\cite{Marciano:2024nwm, Davidson:2002qv, Lu:2019tjj, Kofman:2003nx}, we consider only the decay processes 
$\Sigma_1 \to L H$ (and its CP-conjugate) and the corresponding inverse decay processes $L H \to \Sigma_1$ in our analysis.

We find that, in the considered model, the mass of the fermion triplets lies in the range $10^{11}\text{--}10^{12}\,\mathrm{GeV}$. This mass scale satisfies the Davidson-Ibarra bound required for successful generation of the baryon asymmetry, which requires $M_{\Sigma_1} \gtrsim 10^{9}\,\mathrm{GeV}$~\cite{Davidson:2002qv}. In the type-III seesaw framework, the heavy states correspond to hypercharge-neutral fermion triplets carrying $\mathrm{SU}(2)_L$ gauge interactions. As a consequence, gauge-mediated scattering processes impose a stronger lower bound on the triplet mass, namely $M_{\Sigma_1} \gtrsim 3 \times 10^{10}\,\mathrm{GeV}$~\cite{Vatsyayan:2022rth, Hambye:2003rt}. The mass scale predicted in our model comfortably satisfies both of these constraints. The leptogenesis analysis is carried out at the parameter sets corresponding to the minimum of the $\chi^2$ function, which are taken as benchmark points for each scenario, shown in Table~\ref{tab:lepto_fixed}.

\begin{figure*}[t]
\centering
\includegraphics[width=0.45\linewidth]{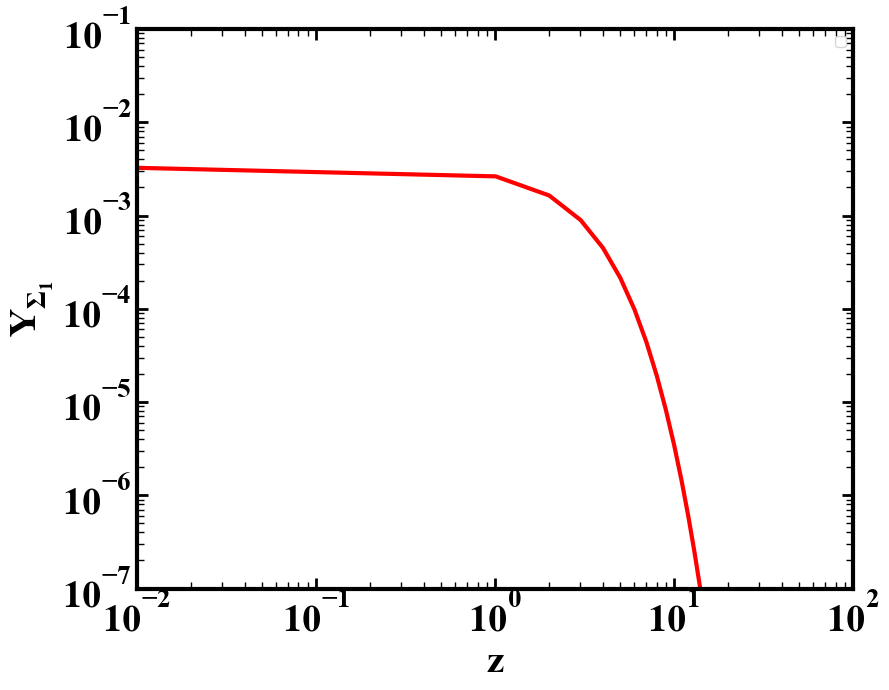}
\includegraphics[width=0.45\linewidth]{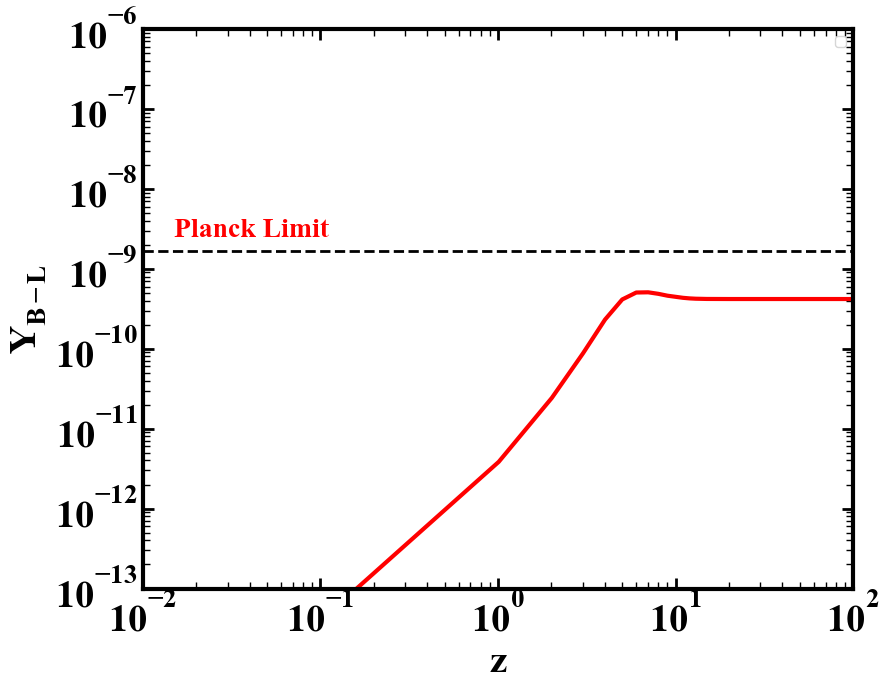}
\caption{Evolution of comoving number density of $\Sigma_1$(Left) and B-L asymmetry (right) as a function of $z = M_{\Sigma_1}/T$ are shown for nearby fixed point $\tau = \frac{3+i}{2}$.}
\label{fig_lepto_3}
\end{figure*}

\begin{figure*}[t]
\centering
\includegraphics[width=0.45\linewidth]{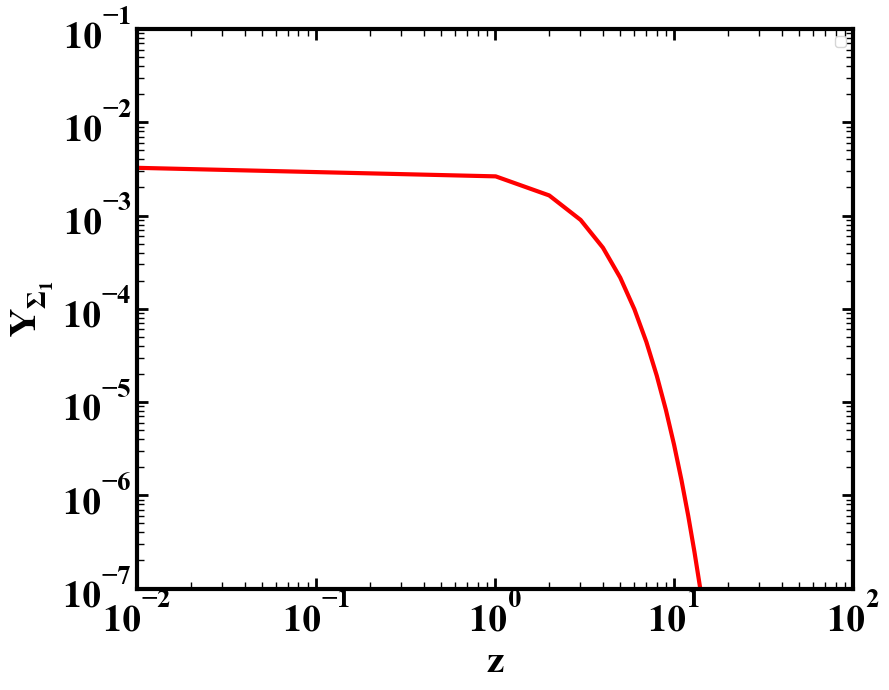}
\includegraphics[width=0.45\linewidth]{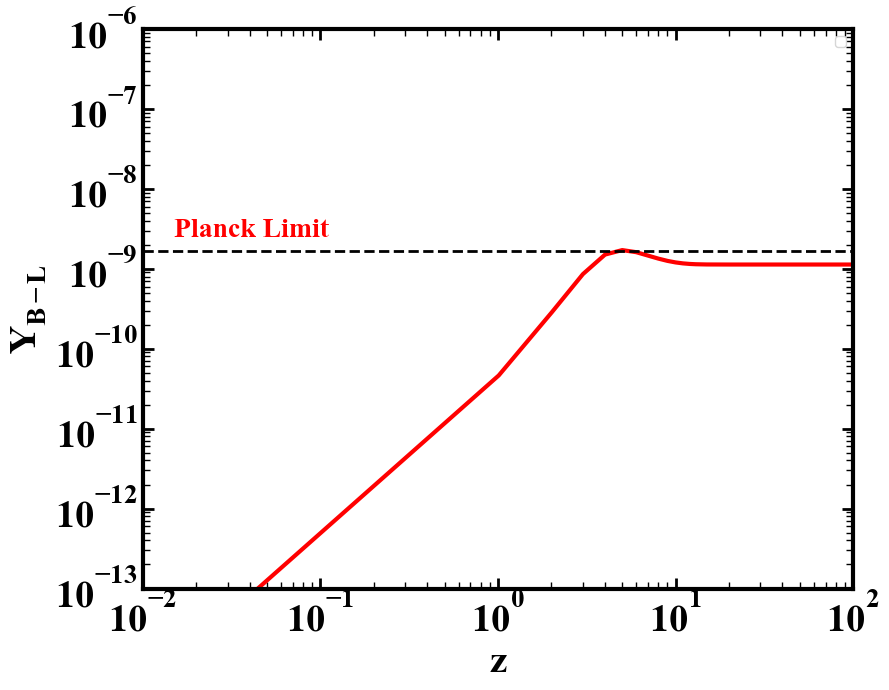}
\caption{Evolution of comoving number density of $\Sigma_1$(Left) and B-L asymmetry (right) as a function of $z = M_{\Sigma_1}/T$ are shown for nearby fixed point $\tau = 1$.}
\label{fig_lepto_7}
\end{figure*}

\begin{figure*}[t]
\centering
\includegraphics[width=0.45\linewidth]{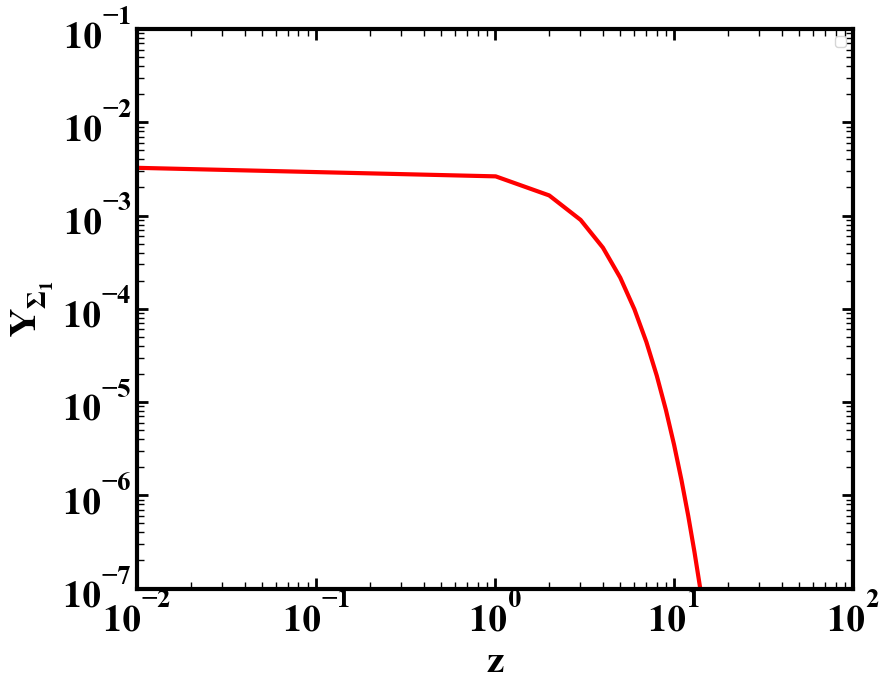}
\includegraphics[width=0.45\linewidth]{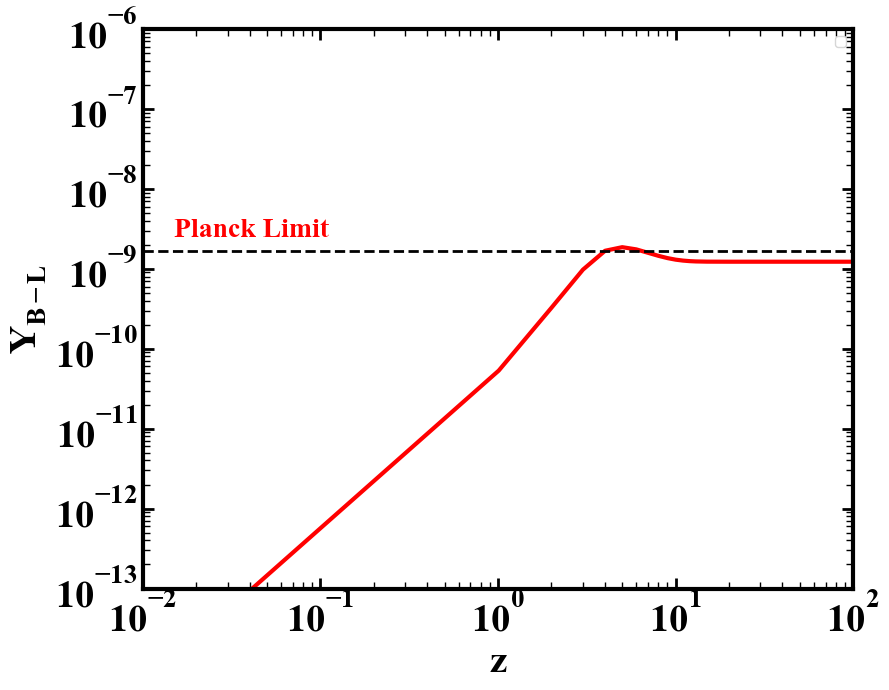}
\caption{Evolution of comoving number density of $\Sigma_1$(Left) and B-L asymmetry (right) as a function of $z = M_{\Sigma_1}/T$ are shown for nearby fixed point $\tau = -1$.}
\label{fig_lepto_8}
\end{figure*}

\noindent We analyze the thermal evolution of the comoving number density of the lightest fermion triplet, $Y_{\Sigma_1}$, together with the generated $B-L$ asymmetry, $Y_{B-L}$, as functions of the dimensionless variable $z=M_{\Sigma_1}/T$ for different values of the modulus $\tau$. For the nearby modular fixed point $\tau=\frac{3+i}{2}$, the evolution of $Y_{\Sigma_1}$ and $Y_{B-L}$ is shown in Fig.~\ref{fig_lepto_3} (Left) and Fig.~\ref{fig_lepto_3} (Right), respectively. We also study the thermal history at $\tau=1$, where the corresponding behavior of $Y_{\Sigma_1}$ and $Y_{B-L}$ is presented in Fig.~\ref{fig_lepto_7} (Left) and Fig.~\ref{fig_lepto_7} (Right). Similarly, the results for $\tau=-1$ are displayed in Fig.~\ref{fig_lepto_8} (Left) for $Y_{\Sigma_1}$ and Fig.~\ref{fig_lepto_8} (Right) for $Y_{B-L}$. The parameter set corresponding to the nearby fixed points used in the leptogenesis analysis, taken as a benchmark scenario are given in Table~\ref{tab:lepto_fixed}. At early times, when $z \ll 1$, the fermion triplet abundance closely tracks its equilibrium value, $Y_{\Sigma_1} \simeq Y_{\Sigma_1}^{\rm eq}$, due to the efficiency of gauge-mediated scattering processes, denoted by $\gamma_A$. These interactions keep the triplet in thermal equilibrium until temperatures around $z \sim \mathcal{O}(1)$. As the Universe cools and $z$ increases beyond unity, the gauge scattering rate becomes smaller than the Hubble expansion rate, causing the triplet abundance to depart from equilibrium. At later times, for $z \gg 10$, the comoving number density of $\Sigma_1$ is exponentially suppressed, as expected from non-relativistic freeze-out.

\begin{table*}[t]
    \centering
    \caption{Benchmark points for baryon asymmetry generation corresponding to the best-fit neutrino oscillation parameters.}
    \begin{tabular}{|c|ccccc|}
        \hline
         Fixed Point & $\chi^2_{min}$ & $M_{\Sigma_1}$ & $\epsilon_{CP}$ & $\rm Re[\tau]$ & $\rm Im[\tau]$ \\
        \hline
        $\frac{3+i\sqrt{2}}{2}$ & $0.17$&$5.46 \times 10^{11} \rm GeV$ & $-0.0000376$ & $1.389$ & $0.536$   \\
        \hline
        $1$ & $0.14$&$2.75 \times 10^{12} \rm GeV$ & $-0.000118$ & $0.958$ & $0.0354$   \\
        \hline
        $-1$ & $0.20$&$3.01 \times 10^{12} \rm GeV$ &$0.000125$ & $-0.958$ & $0.0351$   \\
        \hline 
    \end{tabular}
    \label{tab:lepto_fixed}
\end{table*}

The $B-L$ asymmetry is initially zero and is produced predominantly during the interval $z \sim 1$-$10$. In this temperature range, the decay rate of $\Sigma_1$ into lepton-Higgs states exceeds that of inverse decays and other washout processes. Owing to the CP-violating character of these decays, encoded in the parameter $\varepsilon_{\rm CP}$, a nonvanishing lepton asymmetry is generated. As the Universe further cools and $z$ increases well beyond 10, inverse decay and gauge-mediated scattering processes become inefficient and effectively switch off. Consequently, washout effects freeze out, and the $B-L$ asymmetry settles to a constant value, which constitutes the final asymmetry.

In the relativistic regime, corresponding to $z \ll 1$, the abundance of the fermion triplet remains very close to its equilibrium value, $Y_{\Sigma_1} \approx Y_{\Sigma_1}^{\rm eq}$. This behavior is driven by efficient gauge-mediated scattering processes, denoted by $\gamma_A$, which are sufficiently rapid to keep the triplet in thermal equilibrium up to temperatures of order $T \sim M_{\Sigma_1}$, i.e.\ $z \sim \mathcal{O}(1)$. As the temperature drops further and $z$ becomes larger than unity, the gauge scattering rate falls below the Hubble expansion rate, and the triplet abundance gradually departs from equilibrium. At late times, for $z \gg 10$, the fermion triplet becomes non-relativistic and its comoving number density is exponentially suppressed, as expected from standard freeze-out dynamics. The $B-L$ asymmetry is initially zero and starts to be generated once the system enters the out-of-equilibrium regime, typically for $z$ in the range $1$-$10$. During this phase, the CP-violating decays of $\Sigma_1$ into lepton-Higgs final states dominate over inverse decays and washout effects. The resulting asymmetry is governed by the CP-violating parameter $\varepsilon_{\rm CP}$, which leads to a net lepton asymmetry. At later stages, when $z \gg 10$, inverse decay and gauge scattering processes become inefficient, washout effects freeze out, and the $B-L$ asymmetry stabilizes at a constant final value. 

The fermion triplet, an $SU(2)_L$ multiplet with hypercharge $Y=0$, provides a well-motivated extension of the Standard Model through its role in the Type III seesaw mechanism. In this framework, the triplet couples to the Standard Model lepton doublets and the Higgs field via Yukawa interactions, generating light neutrino masses after electroweak symmetry breaking according to the seesaw relation
\begin{equation}
m_\nu \sim \frac{Y_\Sigma^2\, v^2}{M_\Sigma},
\end{equation}
where $Y_\Sigma$ denotes the Yukawa coupling, $v$ is the Higgs vacuum expectation value, and $M_\Sigma$ is the mass of the fermion triplet.

Despite their central role in the generation of the baryon asymmetry, the fermion triplets predicted in this framework have masses far above the energy reach of current and near-future collider experiments. With $M_{\Sigma_1}$ lying at the scale $10^{11}$-$10^{12}\,\mathrm{GeV}$, direct production of these states at the LHC or proposed high-energy colliders is not feasible, and their effects cannot be probed through low-energy collider observables. Consequently, the leptogenesis mechanism discussed here remains experimentally inaccessible through direct searches. Nevertheless, the existence of such heavy triplet states has profound implications for the early Universe, as their out-of-equilibrium, CP-violating decays provide a natural explanation for the observed baryon asymmetry. In this sense, cosmological observations offer an indirect probe of physics at energy scales far beyond those attainable in laboratory experiments, highlighting the complementary role of early-Universe dynamics in exploring the origin of matter.

\section{Conclusions}
\label{section6}

In this work, we have studied a Type-III seesaw framework incorporating an $\mathrm{SU}(2)_L$ fermion triplet within the context of modular symmetry at its fixed points. The model is constructed using non-holomorphic modular symmetry, in which the Yukawa couplings take a polyharmonic form. Three independent Yukawa couplings with modular weights $0$, $-2$, and $-4$ are present in this model. We have performed $\chi^2$ analysis in the vicinity of the modular fixed points. To explore deviations from the exact fixed points, we scanned the parameter space by deforming the modulus as $\tau \rightarrow \tau(1 + \epsilon e^{i\phi})$, where $\epsilon \in (0,0.1)$ and $\phi \in (-\pi,\pi)$. Within this nearby region, we identified four fixed points that remain consistent with the current neutrino oscillation data. Among them, the fixed point $\tau = \frac{3 + i}{2}$ yields the best fit, with a minimum chi-square value of $\chi^2_{\min} = 0.17$. The corresponding values of the relevant model parameters at this best-fit point are summarized in Table~\ref{tab:neutrino_params}. For the nearby region of fixed point~1, the chi-square minimization yields a minimum value of $\chi^2_{\min} = 0.14$. The corresponding values of the relevant model parameters at this best-fit point are listed in Table~\ref{tab:neutrino_params} . For the nearby region of fixed point~$-1$, the chi-square minimization yields a minimum value of $\chi^2_{\min} = 0.20$. The corresponding values of the relevant model parameters at this best-fit point are presented in Table~\ref{tab:neutrino_params}. In this case as well, the predicted sum of neutrino masses is consistent with the current cosmological bounds. We have also calculated the comoving number density of the lightest right-handed fermion, for which definite predictions are obtained within the model. The Davidson-Ibarra bound corresponding to the lightest neutrino mass is satisfied, and the model provides a consistent framework for the generation of the baryon asymmetry of the Universe through thermal leptogenesis. We find that for the exact modular fixed points, the generated $B-L$ asymmetry agrees well with the observed experimental value. The nearby fixed points also lead to a $B-L$ asymmetry within the allowed experimental range, although they typically correspond to a higher leptogenesis scale. Due to this high energy scale, direct verification at present collider experiments is not feasible. However, such scenarios can be probed indirectly through precision measurements in neutrino oscillation experiments, searches for leptonic CP violation, and cosmological observations sensitive to the thermal history of the early Universe. These studies may provide important tests of the model and its predictions.

\section*{Acknowledgments}
Priya and B. C. Chauhan wants to acknowledge the IUCAA for providing the HPC facility to carry out this work.

\bibliographystyle{JHEP}
\bibliography{zzbib}
\end{document}